\documentclass[12pt]{article}
\usepackage{epsfig,amsmath,amsfonts,amssymb,amstext,afterpage,psfrag,
}
\usepackage{slashed}
\usepackage{graphicx}
\usepackage[usenames,dvipsnames]{color}
\usepackage[square,comma,sort&compress,numbers]{natbib}
\allowdisplaybreaks
\long\def\symbolfootnote[#1]#2{\begingroup%
\def\thefootnote{\fnsymbol{footnote}}\footnote[#1]{#2}\endgroup}

\def\spose#1{\hbox to 0pt{#1\hss}}
\def\lsim{\mathrel{\spose{\lower 3pt\hbox{$\mathchar"218$}}
 \raise 2.0pt\hbox{$\mathchar"13C$}}}
\def\gsim{\mathrel{\spose{\lower 3pt\hbox{$\mathchar"218$}}
 \raise 2.0pt\hbox{$\mathchar"13E$}}}

\catcode`@=11
\def\@citex[#1]#2{%
  \if@filesw\immediate\write\@auxout{\string\citation{#2}}\fi
  \def\@citea{}\@cite{\@for\@citeb:=#2\do
    {\@citea\def\@citea{,\penalty\@m}\@ifundefined
      {b@\@citeb}{{\bf ?}\@warning
{Citation `\@citeb' on page \thepage \space undefined}}%
      \hbox{\csname b@\@citeb\endcsname}}}{#1}}
\def\citer{\@ifnextchar [{\@tempswatrue\@citexr}{\@tempswafalse\@citexr[]}}
  \def\@citexr[#1]#2{%
    \if@filesw\immediate\write\@auxout{\string\citation{#2}}\fi
    \def\@citea{}\@cite{\@for\@citeb:=#2\do
      {\@citea\def\@citea{--\penalty\@m}\@ifundefined
{b@\@citeb}{{\bf ?}\@warning
{Citation `\@citeb' on page \thepage \space undefined}}%
\hbox{\csname b@\@citeb\endcsname}}}{#1}}

\begin{document}

\begin{titlepage}

\begin{flushright}
{\small
LMU-ASC~22/19\\
SI-HEP-2019-05\\
QFET-2019-04\\
}
\end{flushright}

\vspace{0.5cm}
\begin{center}
{\Large\bf \boldmath
Chiral effective theories with a light scalar\\ at one loop
\unboldmath}
\end{center}

\vspace{0.5cm}
\begin{center}
{\sc Oscar Cat\`a$^{a}$ and Christoph M\"uller$^b$}
\end{center}

\vspace*{0.4cm}

\begin{center}
$^a$Theoretische Physik 1, Universit\"at Siegen,\\
Walter-Flex-Stra\ss e 3, D-57068 Siegen, Germany\\
\vspace*{0.2cm}
$^b$Ludwig-Maximilians-Universit\"at M\"unchen, Fakult\"at f\"ur Physik,\\
Arnold Sommerfeld Center for Theoretical Physics,
D-80333 M\"unchen, Germany\\
\end{center}

\vspace{1.5cm}
\begin{abstract}
\vspace{0.2cm}\noindent
There are indications that some theories with spontaneous symmetry breaking also feature a light scalar in their spectrum, with a mass comparable to the one of the Goldstone modes. In this paper, we perform the one-loop renormalization of a theory of Goldstone modes invariant under a chiral $SU(n)\times SU(n)$ symmetry group coupled to a generic scalar singlet. We employ the background field method, together with the heat kernel expansion, to get an expression for the effective action at one loop and single out the anomalous dimensions, which can be read off from the second Seeley-DeWitt coefficient. As a relevant application, we use our master formula to renormalize chiral-scale perturbation theory, an alternative to $SU(3)$ chiral perturbation theory where the $f_0(500)$ meson is interpreted as a dilaton. Based on our results, we briefly discuss strategies to test and discern both effective field theories using lattice simulations.  
\end{abstract}

\vfill

\end{titlepage}

\section{Introduction}
\label{sec:1}
There is a substantial number of spontaneously broken gauge theories whose particle spectrum contains, besides the expected Goldstone modes, a light scalar with a comparable mass. Most of these examples come from lattice studies~\cite{Appelquist:2018yqe,Appelquist:2018tyt,Aoki:2014oha,Aoki:2013zsa,Fodor:2014pqa}, where a large spectrum of quantum field theories can be simulated. The reason behind the lightness of the scalar is in general unclear, though in some cases the vicinity of a fixed point is used as an argument for a dilatonic interpretation. However, in other cases no obvious symmetry argument is found and, similar to the Higgs mass in the Standard Model, there is no explanation for the existence of a light mass scale.

The exploration of gauge theories of spontaneous symmetry breaking with light scalar modes is important in order to gain some more insight into quantum field theories at low energies, especially in theories with strongly-coupled dynamics. In some cases, additionally, there are strong motivations from the phenomenological side. For instance, models of dynamical electroweak symmetry breaking with a natural Higgs boson are still viable scenarios that should be explored~\cite{Fodor:2014pqa}. In these settings, the Higgs boson is most commonly associated with the existence of an infrared fixed point. Depending on how scale symmetry is realized at the fixed point, the light scalar is a scalon or a dilaton. Walking technicolor~\cite{Appelquist:2010gy} and crawling technicolor~\cite{Cata:2018wzl} are examples of how these ideas are, respectively, implemented.

The remarkable thing about strongly-coupled theories with spontaneous symmetry breaking is that a mass gap is created and the dynamics of the Goldstone modes can be described with a low-energy effective field theory (EFT), where the known (hidden) symmetry pattern constrains the form of the effective operators. The canonical example is the theory of the strong interactions at low energies, where chiral symmetry breaking generates the pion multiplet, whose interactions are described with chiral perturbation theory (ChPT). 

Building effective theories of Goldstone bosons can be done with the CCWZ formalism~\cite{Callan:1969sn}, and extending them with a generic light scalar field is rather straightforward. As a result, one can describe the low-energy dynamics of a large class of theories in a systematic way. Following these methods, low-energy theories with Goldstones and a light scalar are receiving increasing attention~\cite{Soto:2011ap,Golterman:2016lsd,Hansen:2016fri,Hansen:2018gck}. 

In this paper, we perform the one-loop renormalization of theories of spontaneously broken $SU(n)\times SU(n)$ enlarged with a singlet scalar. EFTs of Goldstone modes are organized as expansions in the number of loops. As such, divergences at a certain loop order can always be absorbed by a finite number of counterterms. The EFT is nonrenormalizable as a whole but is renormalizable order by order. Here we will concentrate on the counterterms needed to absorb the one-loop divergences and compute their anomalous dimensions. In order to make the calculation more efficient, we will employ well-known functional methods. The background field method (see e.g.~\cite{Abbott:1981ke}) is especially suited to keep the symmetries of the calculation at all times explicit and the heat kernel expansion~\cite{DeWitt:1967yk} of the fluctuation determinant provides a very efficient method to single out the divergences. Since we will work in dimensional regularization, the divergences will come out as the residues of the $1/\epsilon$ poles.

The resulting master formula for the anomalous dimensions applies to chiral $SU(n)\times SU(n)$ effective theories with a generic scalar. In order to validate it, we have tested it with the $O(4)$ linear sigma model and $SU(3)$ chiral perturbation theory, whose one-loop renormalization is known. As a relevant new application, we have renormalized chiral-scale perturbation theory (ChSPT)~\cite{Crewther:2012wd,Crewther:2013vea,Crewther:2015dpa}. This low-energy EFT stands as an alternative to conventional $SU(3)$ ChPT, where the $f_0(500)$ meson is interpreted as a dilaton and thus included as a light degree of freedom. With our results, ChSPT can be tested at the one-loop level. Some comments on how to do that with lattice simulations are given.    

This paper is organized as follows: in Section~\ref{sec:2}, we summarize the basics of a general chiral $SU(n)$ effective theory with a light scalar. We discuss the leading-order Lagrangian, the associated power counting and finally list down the basis of next-to-leading order operators. Section~\ref{sec:3} is devoted to the computation of the divergent pieces of the one-loop renormalization. Master expressions for the anomalous dimensions are provided. In Section~\ref{sec:4}, we apply the previous results to ChSPT and draw some comments on how it can be compared to conventional $SU(3)$ ChPT at one loop. Concluding remarks are given in Section~\ref{sec:5}.


\section{Chiral effective theories with a light scalar}
\label{sec:2}

The construction of effective field theories for systems with spontaneous symmetry breaking relies on the CCWZ formalism~\cite{Callan:1969sn}, which parametrizes the Goldstone modes as fluctuations over the vacuum manifold. We will restrict our attention to theories with a $SU(n)_L\times SU(n)_R\to SU(n)_V$ dynamical symmetry breaking pattern, such as theories of fundamental fermions with light or vanishing masses that undergo fermion condensation, an example of which is QCD. In this case, the CCWZ formalism gets simplified and the Goldstone bosons, which we will denote generically as $\pi^a$, can be collected in the unitary matrix
\begin{align}
U(x)&=e^{2i\pi^a t^a/f_\pi}\,,
\end{align} 
where $f_\pi$ is the Goldstone decay constant and $t^a$ are the generators of the broken $SU(n)$, normalized such that ${\mathrm{tr}}[t^at^b]=\tfrac{1}{2}\delta^{ab}$. The matrix $U(x)$ transforms as a bifundamental under the chiral group, i.e.
\begin{align}
U(x)\to g_L U(x) g_R^{-1}\,.
\end{align}
Goldstone dynamics is the EFT built from $U$ and its derivatives. It is convenient to define a covariant derivative with generic left- and right-handed spurion gauge fields, such that
\begin{align}
D_{\mu}U&=\partial_{\mu}U+il_{\mu}U-iUr_{\mu}\,,
\end{align}
with 
\begin{align}
l_{\mu}\to g_L l_{\mu} g_L^{-1}-ig_L\partial_{\mu}g_L^{-1};\qquad r_{\mu}\to g_R r_{\mu} g_R^{-1}-ig_R\partial_{\mu}g_R^{-1}\,.
\end{align}
If one also allows for explicit symmetry breaking in the form of fermion mass terms, a spurion field $\chi$ transforming as $\chi\to g_L \chi g_R^{-1}$ is the most efficient way to systematically write down the operators of explicit breaking. 

In the following, we will use the alternative, more compact notation, given by the field
\begin{align}
u(x)=e^{i\pi^a t^a/f_\pi};\qquad u(x)\to g_L u(x) h^{\dagger}=h u(x) g_R^{-1},\qquad h\in SU(n)_V\,,
\end{align}
out of which one has the building blocks 
\begin{align}
u_{\mu}&=iuD_{\mu}U^{\dagger}u\nonumber\\
\chi_{\pm}&=u^{\dagger} \chi u^{\dagger}\pm u\chi^{\dagger}u\nonumber\\
f_{\pm}^{\mu\nu}&=uf_R^{\mu\nu}u^{\dagger}\pm u^{\dagger}f_L^{\mu\nu}u\,,
\end{align}
where $f_{R}^{\mu\nu}=\partial^{\mu}r^{\nu}-\partial^{\nu}r^{\mu}-i[r^{\mu},r^{\nu}]$ and similarly for $f_{L}^{\mu\nu}$. All the matrices defined above transform in the adjoint of $SU(n)_V$, $X\to h X h^{\dagger}$, which makes it easier to build invariant operators. 

The leading-order Lagrangian is constructed such that it contains the Goldstone kinetic term, i.e.
\begin{align}
{\cal{L}}_2&=\frac{f_\pi^2}{4}\langle u_{\mu}u^{\mu}+\chi_+\rangle\,,
\end{align}
where $\langle...\rangle$ denotes the trace. If one assumes that parity is a good symmetry of the theory, as we will do, the previous operators exhaust the leading order in the EFT.

Extending the previous EFT with an arbitrary singlet scalar $\phi$ is rather straightforward. Based on symmetry arguments, the most general leading-order Lagrangian is
\begin{align}\label{LO}
{\cal{L}}_2&=\frac{f_\pi^2}{4}\langle u_{\mu}u^{\mu}\rangle f(\phi)+\frac{f_\pi^2}{4}\langle\chi_+\rangle g(\phi)+\frac{1}{2}\partial_{\mu}\phi\partial^{\mu}\phi h(\phi)-V(\phi)\,,
\end{align}
where the form of the dimensionless scalar functions is undetermined. Expanding them in polynomial form, they can be expressed as
\begin{align}\label{functions}
f(\phi)&=1+\sum_{j=1}^{\infty} a_j \left(\frac{\phi}{f_{\phi}}\right)^j; \quad g(\phi)=1+\sum_{j=1}^{\infty} b_j \left(\frac{\phi}{f_{\phi}}\right)^j;\nonumber\\
h(\phi)&=1+\sum_{j=1}^{\infty} d_j \left(\frac{\phi}{f_{\phi}}\right)^j; \quad V(\phi)=m_{\phi}^2f_{\phi}^2\sum_{j=2}^{\infty} c_j \left(\frac{\phi}{f_{\phi}}\right)^j\,,
\end{align}
with the coefficients being model-dependent. The scale $f_{\phi}$ is introduced only to make the coefficients above dimensionless; it does not need to be a dynamical scale. Unlike for the chiral Goldstones, there is in general not a unique coupling for the scalar interactions. The number of independent couplings will be determined by the symmetries of the scalar sector.

Eq.~(\ref{LO}) contains an unlimited number of interaction vertices. This is a generic characteristic of scalar theories coupled to Goldstones. It is important to stress that this consideration applies regardless of whether the scalar is itself a Goldstone mode or not. This can be seen by deriving the power-counting formula linked to eq.~(\ref{LO}). 

Let us define the vertices of the theory as $V_{ij}^{(2)}$ and $V_{ij}^{(0)}$, where $i,j$ denote the number of Goldstone and scalar lines attached to the vertex and the superscript counts the number of derivatives. From topological considerations, the naive degree of divergence of any diagram built from eq.~(\ref{LO}) is
\begin{align}\label{top1}
d&=4\ell-2p_{\phi}-2p_{\pi}+2\sum_{i,j}V_{ij}^{(2)}\,,
\end{align}
where $p_j$ are the number of propagators of each field. The number of loops $\ell$ is given in turn by
\begin{align}\label{top2}
\ell&=p_{\pi}+p_{\phi}-\sum_{i,j}V_{ij}^{(2)}-\sum_{i,j}V_{ij}^{(0)}+1\,.
\end{align} 
Combining the previous equations two equations, one finds the compact result:
\begin{align}\label{top3}
d&=2\ell+2-2\sum_{i,j}V_{ij}^{(0)}\,.
\end{align} 
The first observation to make concerning eqs.~(\ref{top1})-(\ref{top3}) is that they do not depend on the number of external legs. In other words, at each loop level, one can have an arbitrary number of external Goldstone and scalar fields. This justifies the expressions for the scalar functions in eq.~(\ref{functions}).

The second observation is that the sum over non-derivative interactions can be traded for the total number of $\chi$ and $c_j$ insertions. This suggests to write the loop counting as
\begin{align}\label{loop}
2\ell+2=d+2n_{\chi}+2n_{c_j}\,,
\end{align}  
where $n_i$ is the number of parameter insertions. 

Theories with scalars and Goldstone modes are thus organized as loop expansions, where both the number of derivatives and the number of insertions of the potential parameters count (see the discussion in~\cite{Buchalla:2012qq,Buchalla:2013eza,Buchalla:2013rka,Buchalla:2016sop}). In view of eq.~(\ref{loop}), one can define the chiral dimension for an object $A$ (a field or a coupling), denoted by $[A]_{\chi}$, as~\cite{Buchalla:2013eza,Urech:1994hd}       
\begin{align}\label{dc}
[U;\phi]_{\chi}&=0; & [\partial_{\mu};u_{\mu}]_{\chi}&=1; &[c_j;\chi_{\pm}]_{\chi}&=2\,.
\end{align}
This prescription has the advantage that it allows to interpret the diagrammatic degree of divergence at the operator level. With eq.~(\ref{dc}), one can catalog the different operators in the chiral expansion at any order. The LO Lagrangian corresponds to $\ell=0$, which can be achieved with the combinations $(d,n_{\chi},n_{c_j})=\{(2,0,0);(0,1,0);(0,0,1)\}$. This list is exhausted with the operators of eq.~(\ref{LO}), all of them satisfying $[{\cal{O}}_{LO}]_{\chi}=2$.

The NLO operators can now be built from invariant combinations of the building blocks satisfying $[{\cal{O}}_{NLO}]_{\chi}=4$. This corresponds to the combinations $(d,n_{\chi},n_{c_j})=\{(4,0,0);(0,2,0);(2,1,0)\}$. Notice that the classes $(2,0,1)$ and $(0,0,2)$ do not generate new independent operators but merely provide NLO corrections to existing LO operators. Using parity conservation and eliminating redundancies in favor of operators with the minimal number of derivatives, one arrives at the following basis of NLO operators:
\begin{align}\label{NLO}
{\cal{L}}_4&=L_0(\phi)\langle u_{\mu}u_{\nu}u^{\mu}u^{\nu}\rangle+L_1(\phi)\langle u_{\mu}u^{\mu}\rangle^2+L_2(\phi)\langle u^{\mu}u^{\nu}\rangle\langle u_{\mu}u_{\nu}\rangle+L_3(\phi)\langle u_{\mu}u^{\mu}u_{\nu}u^{\nu}\rangle\nonumber\\
&+L_4(\phi)\langle u_{\mu}u^{\mu}\rangle\langle\chi_+\rangle+L_5(\phi)\langle u_{\mu}u^{\mu}\chi_+\rangle+L_6(\phi)\langle\chi_+\rangle^2+L_7(\phi)\langle\chi_-\rangle^2+\frac{1}{2}L_8(\phi)\langle\chi_+^2+\chi_-^2\rangle\nonumber\\
&-iL_9(\phi)\langle f_+^{\mu\nu}u_{\mu}u_{\nu}\rangle+\frac{1}{4}L_{10}(\phi)\langle (f_+^{\mu\nu})^2-(f_-^{\mu\nu})^2\rangle+H_1(\phi)\langle (f_R^{\mu\nu})^2+(f_L^{\mu\nu})^2\rangle+H_2(\phi)\langle\chi^{\dagger}\chi\rangle\nonumber\\
&+G_1(\phi)\partial_{\mu}\phi\partial^{\mu}\phi\partial_{\nu}\phi\partial^{\nu}\phi+G_2(\phi)\langle u_{\mu}u^{\mu}\rangle \partial_{\nu}\phi\partial^{\nu}\phi+G_3(\phi)\langle u_{\mu}u_{\nu}\rangle \partial^{\mu}\phi\partial^{\nu}\phi\nonumber\\
&+G_4(\phi)\langle\chi_+\rangle \partial_{\mu}\phi\partial^{\mu}\phi+\frac{i}{2}G_5(\phi)\langle u_{\mu}\chi_-\rangle\partial^{\mu}\phi+G_6(\phi)\langle u_{\mu}f_-^{\mu\nu}\rangle\partial_{\nu}\phi\,,
\end{align} 
where the scalar functions $L_j(\phi)$ are defined as
\begin{align}\label{expansion}
L_j(\phi)&=L_j+\sum_k l_j^{(k)}\left(\frac{\phi}{f_{\phi}}\right)^k\,,
\end{align}
and similarly for $H_j(\phi)$ and $G_j(\phi)$. 

One of the nice features of using chiral dimensions is that they exhaust the list of operators at a given order, NLO in this case. Depending on the scalar theory at hand, some of the operators above will act as counterterms, while some others will bring finite contributions to the NLO amplitudes of the theory.

\section{One-loop renormalization}
\label{sec:3}
The one-loop divergences from eq.~(\ref{LO}) will be absorbed by some of the NLO operators listed in eq.~(\ref{NLO}), whose coefficients will be renormalized. In dimensional regularization with a modified ${\overline{MS}}$ scheme, one has
\begin{align}\label{anom}
L_i(\phi)=L_i^r(\phi;\mu)-\Gamma_i(\phi)\lambda;\quad H_i(\phi)=H_i^r(\phi;\mu)-\Delta_i(\phi)\lambda;\quad G_i(\phi)=G_i^r(\phi;\mu)-\widehat\Gamma_i(\phi)\lambda\,,
\end{align}
where $\Gamma_i(\phi)$, $\Delta_i(\phi)$ and $\widehat\Gamma_i(\phi)$ are the anomalous dimensions and
\begin{align}
\lambda=\frac{\mu^{-2\epsilon}}{32\pi^2}\left(\frac{1}{\epsilon}+\log 4\pi-\gamma_E+1\right)\,,
\end{align}
with $\epsilon=2-d/2$ and $\gamma_E$ being the Euler-Mascheroni constant. 

Eqs.~(\ref{anom}), as opposed to what happens with pure Goldstone theories, are functional relations. The anomalous coefficients are defined from the series expansion of the anomalous functions, e.g.
\begin{align}
\Gamma_i(\phi)=\Gamma_i+\sum_k \Gamma_{i}^{(k)}\left(\frac{\phi}{f_{\phi}}\right)^k\,.
\end{align}
$\Gamma_{i}^{(k)}$ is then the anomalous dimension (a $SU(n)$-dependent number) associated to the coefficient $l_{i}^{(k)}$ of the corresponding expansion of $L_i(\phi)$. 
 
\subsection{Determination of the anomalous dimensions}

The anomalous dimensions can be obtained most efficiently by using the background field method (see e.g.~\cite{Abbott:1981ke,Gasser:1983yg}). For a generic Lagrangian depending only on a set of scalar fields and their first derivatives, one splits the fields into classical background fields $\phi_j$ and quantum fluctuations $\varphi_j$, where the subindex emphasizes the vectorial character of these fields, which in $SU(n)$ chiral theories with a scalar have $n^2$ components. 

The one-loop contributions to the effective action can be computed by expanding the Lagrangian to second order in the fluctuations: 
\begin{align}\label{fluc}
{\cal{L}}(\phi+\varphi)={\cal{L}}_0(\phi)+A^i\varphi_i-\frac{1}{2}\varphi_i B^{ij}\varphi_j+{\cal{O}}(\varphi^3)\,.
\end{align}
It is convenient to identify the background fields with the classical fields. Then the linear term above cancels by using the equations of motion. By appropriate field redefinitions and integration by parts, the quadratic term can always be written (up to total derivatives) as
\begin{align}\label{quad}
B=(\partial^{\mu}+\omega^{\mu})(\partial_{\mu}+\omega_{\mu})-\sigma\,,
\end{align} 
where 
\begin{align}\label{symm}
\omega^\mu_{ij}=-\omega^\mu_{ji};\qquad \sigma_{ij}=\sigma_{ji}\,.
\end{align} 
In this form, the operator $B$ is elliptic and leads to a well-defined Gaussian integral in Euclidean space. The anomalous dimensions can be extracted from the heat kernel expansion of the resulting functional determinant, in particular from the so-called second Seeley-DeWitt coefficient. In Minkowski space, this divergent piece is given by~\cite{DeWitt:1967yk,tHooft:1973bhk}
\begin{align}
\Delta{\cal{L}}=\frac{1}{32\pi^2\epsilon}\left[\frac{1}{12}\langle \omega_{\mu\nu}\omega^{\mu\nu}\rangle+\frac{1}{2}\langle \sigma^2\rangle\right]\,,
\end{align}
where $\omega_{\mu\nu}=\partial_\mu \omega_\nu-\partial_\nu\omega_\mu+[\omega_\mu,\omega_\nu]$ and the trace is over all internal indices. Given our conventions of eqs.~(\ref{anom}), the anomalous dimensions are simply the coefficients in front of the operators generated by $\tfrac{1}{12}\langle\omega_{\mu\nu}\omega^ {\mu\nu}\rangle+\tfrac{1}{2}\langle\sigma^2\rangle$.

In order to apply the formalism to eq.~(\ref{LO}), we need to define how to parametrize the fluctuations around background fields. The final result is of course independent of the parametrization, but a good choice simplifies the algebra in the intermediate steps. We will adopt the convention
\begin{align}
U=e^{i\Delta}U_0;\qquad \phi=\phi_0+\delta\,.
\end{align}
In order to simplify the notation, from now on we will drop the subindex. $U$ and $\phi$ will henceforth be understood as the background fields. With our conventions, the Goldstone fluctuation is left-handed, $e^{i\Delta}\to g_L e^{i\Delta} g_L^{-1}$, and its covariant derivative is
\begin{align}
D_{\mu}\Delta=\partial_{\mu}\Delta+i[l_{\mu},\Delta]\,.
\end{align}
In order to reduce the number of chiral matrices inside operators, with these conventions it is convenient to define the left-handed chiral derivative 
\begin{align}
\xi_{\mu}\equiv uu_{\mu}u^{\dagger}=iUD_{\mu}U^{\dagger}\,.
\end{align}
The transition from $\xi_{\mu}$ to $u_{\mu}$ is straightforward, given that $\langle \xi_{\mu}\xi_{\nu}\rangle=\langle u_{\mu}u_{\nu}\rangle$.

The expansion of the leading-order Lagrangian at linear order gives the equations of motion for the classical fields:
\begin{align}
\Box \phi&=\frac{f_\pi^2}{4}\langle u_{\mu}u^{\mu}\rangle\frac{f'}{h}-\frac{1}{2}\partial_{\mu}\phi\partial^{\mu}\phi\frac{h'}{h}+\frac{f_\pi^2}{4}\langle\chi_+\rangle\frac{g'}{h}-\frac{V'}{h}\nonumber\\
D^{\mu}\xi_{\mu}^a&=-\frac{f'}{f}\xi_{\mu}^a\partial^{\mu}\phi-\frac{i}{2}\langle u\chi_-u^{\dagger}t^a\rangle\frac{g}{f}\,,
\end{align}
where our conventions are such that $\xi_{\mu}^a=\langle \xi_{\mu}t^a\rangle$ and $\Delta^a=\langle\Delta t^a\rangle$. The LO equations of motion will be used at different stages of the computation in order to simplify the algebra and stick to our conventions of having a NLO with the minimal number of derivatives. In EFTs, using the LO equations of motion to perform simplifications at subleading orders is equivalent to a field redefinition~\cite{Georgi:1991ch}, and it is therefore always allowed.
 
With a bit of algebra, one can express the quadratic fluctuations in the canonical form of eqs.~(\ref{fluc})-(\ref{symm}) and thus find the matrix elements that build the Seeley-DeWitt coefficients. In Euclidean space, they read
\begin{align}
(\omega_{\mu})^{00}&=0; & \sigma^{00}&=\frac{V''}{h}-\frac{V'h'}{2h^2}-\frac{f}{h}{\cal{B}}\xi_{\mu}^b\xi^{\mu b}-\frac{{\cal{D}}}{4}\langle\chi_+\rangle\nonumber\\
(\omega_{\mu})^{a0}&=-(\omega_{\mu})^{0a}=\sqrt{a}\xi_{\mu}^a; & \sigma^{a0}&=\frac{{\cal{B}}}{f_{\pi}}\sqrt{\frac{f}{h}}\xi_{\mu}^a\partial^{\mu}\phi-i\frac{{\cal{C}}}{2}\frac{g}{\sqrt{fh}}\langle u\chi_-u^{\dagger} t^a\rangle\nonumber\\
(\omega_{\mu})^{ab}&=-f^{abc}(\xi_{\mu}^c-l_{\mu}^c); &\sigma^{ab}&=a(\delta^{ab}\xi_{\mu}^c\xi^{\mu c}-\xi_{\mu}^a\xi^{\mu b})-f^{aic}f^{bjc}\xi_{\mu}^i\xi^{\mu j}-\frac{{\cal{B}}}{f_{\pi}^2}\delta^{ab}\partial_{\mu}\phi\partial^{\mu}\phi\nonumber\\
&&&+\delta^{ab}\frac{V'f'}{2fh}+\frac{g}{f}\langle t^at^b\chi_+\rangle-\delta^{ab}\frac{b}{4}\frac{g}{f}\langle\chi_+\rangle\,,
\end{align}
where $l_{\mu}^a=2\langle l_{\mu}t^a\rangle$ and $f^{abc}$ are the $SU(n)$ structure constants. For later convenience we have defined the dimensionless combinations
\begin{align}\label{rel1}
a=\frac{f_\pi^2}{4}\frac{f'^2}{fh}; \qquad b=\frac{f_\pi^2}{2}\frac{f'g'}{gh}
\end{align}
and
\begin{align}\label{rel2}
{\cal{B}}&=\frac{f_\pi^2}{4}\left(\frac{f'^2}{f^2}+\frac{f'h'}{fh}-2\frac{f''}{f}\right)\nonumber\\
{\cal{C}}&=f_\pi\left(\frac{1}{2}\frac{f'}{f}-\frac{g'}{g}\right)\nonumber\\
{\cal{D}}&=f_\pi^2\left(\frac{g''}{h}-\frac{1}{2}\frac{g'h'}{h^2}\right)\,.
\end{align}
Assembling all the pieces together, one ends up with the result (in Minkowski space): 
\begin{align}\label{result}
\frac{1}{12}&\langle\omega_{\mu\nu}\omega^{\mu\nu}\rangle+\frac{1}{2}\langle\sigma^2\rangle=\frac{8a+n}{24}\langle u_{\mu}u^{\mu}u_{\nu}u^{\nu}\rangle-\frac{16a-n}{48}\langle u_{\mu}u_{\nu}u^{\mu}u^{\nu}\rangle\nonumber\\
&+\frac{4a^2+3}{24}\langle u_{\mu}u_{\nu}\rangle^2+\frac{1}{4}\left(\frac{f^2}{2h^2}{\cal{B}}^2+\frac{3n^2-10}{6}a^2-na+\frac{1}{4}\right)\langle u_{\mu}u^{\mu}\rangle^2+\frac{2a+n}{8}\frac{g}{f}\langle u_{\mu}u^{\mu}\chi_+\rangle\nonumber\\
&-\frac{1}{4}\left[\frac{f}{h}\frac{{\cal{B}}{\cal{D}}}{2}+\frac{g}{f}\left(\frac{n^2-1}{n}a-\frac{1}{2}\right)-\frac{g}{f}b\left(\frac{n^2-2}{2}a-\frac{n}{2}\right)\right]\langle u_{\mu}u^{\mu}\rangle\langle\chi_+\rangle\nonumber\\
&+\frac{1}{32}\left[{\cal{D}}^2+3\left(\frac{g}{f}\right)^2(1-b)^2+(n-2)\left(\frac{g}{f}\right)^2\left((n+2)b^2-\frac{4n+2}{n}b-\frac{n+2}{n^2}\right)\right]\langle\chi_+\rangle^2\nonumber\\
&+\frac{n^2-4}{16n}\left(\frac{g}{f}\right)^2\langle\chi_+^2\rangle-\frac{1}{8}\frac{g^2}{fh}{\cal{C}}^2\langle\chi_-^2\rangle+\frac{1}{8n}\frac{g^2}{fh}{\cal{C}}^2\langle\chi_-\rangle^2+i\frac{2a-n}{12}\langle f_+^{\mu\nu}u_{\mu}u_{\nu}\rangle\nonumber\\
&+\frac{2a-n}{12}\langle f_{L}^{\mu\nu}Uf_{R\mu\nu}U^{\dagger}\rangle-\frac{2a+n}{24}\langle f_{L\mu\nu}f_L^{\mu\nu}+f_{R\mu\nu}f_R^{\mu\nu}\rangle+\frac{n^2-1}{2f_\pi^4}{\cal{B}}^2(\partial_{\mu}\phi\partial^{\mu}\phi)^2\nonumber\\
&-\frac{1}{f_\pi^2}{\cal{B}}\left(\frac{f}{6h}{\cal{B}}+\frac{n^2-2}{2}a-\frac{n}{2}\right)\langle u_{\mu}u^{\mu}\rangle \partial_{\nu}\phi\partial^{\nu}\phi+\frac{2}{3f_\pi^2}\frac{f}{h}{\cal{B}}^2\langle u_{\mu}u_{\nu}\rangle\partial^{\mu}\phi\partial^{\nu}\phi-\frac{f'}{6h}{\cal{B}}\langle f_-^{\mu\nu}u_{\nu}\rangle\partial_{\mu}\phi\nonumber\\
&+\frac{3}{4f_\pi^2}\frac{g}{f}{\cal{B}}\left(1+(n-2)\frac{2n+1}{3n}-b\frac{n^2-1}{3}\right)\langle\chi_+\rangle\partial_{\mu}\phi\partial^{\mu}\phi+\frac{i}{2f_\pi}\frac{g}{h}{\cal{B}}{\cal{C}}\langle u_{\mu}\chi_-\rangle \partial^{\mu}\phi\,,
\end{align}
where only the NLO structures have been retained. 

In order to obtain the previous result, the following $SU(n)$ relations are useful:
\begin{align}
t^a_{ij}t^a_{kl}&=\frac{1}{2}\delta_{il}\delta_{jk}-\frac{1}{2n}\delta_{ij}\delta_{kl}\nonumber\\
f_{abc}f_{abd}&=n\delta_{cd}\,,
\end{align}
from which one can derive
\begin{align}
f^{aic}f^{bjc}\xi_{\mu}^i\xi^{\mu j}\xi_{\nu}^a\xi^{\nu b}&=-\frac{1}{8}\langle[\xi_{\mu},\xi_{\nu}][\xi^{\mu},\xi^{\nu}]\rangle\nonumber\\
f^{aic}f^{bjc}\xi_{\mu}^i\xi^{\mu j}\langle \chi_+t^at^b\rangle&=\frac{n}{8}\langle\xi_{\mu}\xi^{\mu}\chi_+\rangle+\frac{1}{8}\langle\chi_+\rangle\langle\xi_{\mu}\xi^{\mu}\rangle\nonumber\\
f^{aic}f^{bjc}f^{amd}f^{bkd}\xi_{\mu}^i\xi^{\mu j}\xi_{\nu}^m\xi^{\nu k}&=\frac{1}{4}\langle \xi_{\mu}\xi_{\nu}\rangle^2+\frac{1}{8}\langle\xi_{\mu}\xi^{\mu}\rangle^2+\frac{n}{8}\langle \xi_{\mu}\xi^{\mu}\xi_{\nu}\xi^{\nu}\rangle\nonumber\\
\langle\chi_+t^at^b\rangle^2&=\frac{n^2-4}{8n}\langle\chi_+^2\rangle+\frac{n^2+2}{8n^2}\langle\chi_+\rangle^2\,.
\end{align}
The anomalous dimensions of the operators given in eq.~(\ref{NLO}) can be straightforwardly read off from the coefficients of eq.~(\ref{result}) to yield
\begin{align}
&\Gamma_0^{(n)}=-\frac{a}{3}+\frac{n}{48}; \hspace{5cm}  \Gamma_1^{(n)}=\frac{f^2}{8h^2}{\cal{B}}^2+\frac{3n^2-10}{24}a^2-\frac{n}{4}a+\frac{1}{16}\nonumber\\
&\Gamma_2^{(n)}=\frac{a^2}{6}+\frac{1}{8}; \hspace{5.3cm}  \Gamma_3^{(n)}=\frac{a}{3}+\frac{n}{24}\nonumber\\
&\Gamma_4^{(n)}=-\frac{1}{4}\left[\frac{f}{h}\frac{{\cal{B}}{\cal{D}}}{2}+\frac{g}{f}\left(\frac{n^2-1}{n}a-\frac{1}{2}\right)-\frac{g}{f}b\left(\frac{n^2-2}{2}a-\frac{n}{2}\right)\right];\qquad \Gamma_5^{(n)}=\frac{2a+n}{8}\frac{g}{f}\nonumber\\
&\Gamma_6^{(n)}=\frac{1}{32}\left[{\cal{D}}^2+3\left(\frac{g}{f}\right)^2(1-b)^2+(n-2)\left(\frac{g}{f}\right)^2\left((n+2)b^2-\frac{4n+2}{n}b-\frac{n+2}{n^2}\right)\right]\nonumber\\
&\Gamma_7^{(n)}=\frac{1}{8n}\frac{g^2}{fh}{\cal{C}}^2; \hspace{5cm} \Gamma_8^{(n)}=\frac{n^2-4}{16n}\left(\frac{g}{f}\right)^2-\frac{1}{8}\frac{g^2}{fh}{\cal{C}}^2\nonumber\\
&\Gamma_9^{(n)}=\frac{n}{12}-\frac{a}{6}; \hspace{5.2cm} \Gamma_{10}^{(n)}=-\frac{n}{12}+\frac{a}{6}\nonumber\\
&\widehat\Gamma_{1}^{(n)}=\frac{n^2-1}{2f_\pi^4}{\cal{B}}^2; \hspace{4.8cm} \widehat\Gamma_{2}^{(n)}=-\frac{1}{f_\pi^2}{\cal{B}}\left(\frac{f}{6h}{\cal{B}}+\frac{n^2-2}{2}a-\frac{n}{2}\right)\nonumber\\
&\widehat\Gamma_{3}^{(n)}=\frac{2}{3f_\pi^2}\frac{f}{h}{\cal{B}}^2; \hspace{4.9cm} \widehat\Gamma_{4}^{(n)}=\frac{3}{4f_\pi^2}\frac{g}{f}{\cal{B}}\left(1+(n-2)\frac{2n+1}{3n}-b\frac{n^2-1}{3}\right)\nonumber\\
&\widehat\Gamma_{5}^{(n)}=\frac{1}{f_\pi}\frac{g}{h}{\cal{B}}{\cal{C}}; \hspace{4.9cm} \widehat\Gamma_{6}^{(n)}=-\frac{1}{6}\frac{f'}{h}{\cal{B}}\nonumber\\
&{\Delta}_{1}^{(n)}=-\frac{a}{12}-\frac{n}{24}; \hspace{4.4cm} {\Delta}_{2}^{(n)}=\frac{n^2-4}{8n}\left(\frac{g}{f}\right)^2+\frac{1}{4}\frac{g^2}{fh}{\cal{C}}^2\,.
\end{align} 
Notice from the previous result that the renormalization of the NLO operators is insensitive to the form of the scalar potential. However, the scalar potential plays a paramount role in the renormalization of the leading-order Lagrangian. The functions $f(\phi)$, $g(\phi)$, $h(\phi)$ and $V(\phi)$ have to be renormalized with the following anomalous dimensions: 
\begin{align}
\Gamma_f&=\frac{2{\cal{B}}}{f_{\pi}^2}\frac{f}{h}\left(\frac{V''}{h}-\frac{V'h'}{2h^2}\right)+\frac{n}{f_{\pi}^2}\frac{V'f'}{fh}-(n^2-2)\frac{a}{f_{\pi}^2}\frac{V'f'}{fh}\nonumber\\
\Gamma_g&=-\frac{{\cal{D}}}{f_{\pi}^2}\left(\frac{V''}{h}-\frac{V'h'}{2h^2}\right)-\frac{(n^2-1)(nb-2)}{2nf_{\pi}^2}\frac{g}{f}\frac{V'f'}{fh}\nonumber\\
\Gamma_h&=(n^2-1)\frac{{\cal{B}}}{f_{\pi}^2}\frac{V'f'}{fh}\nonumber\\
\Gamma_V&=\frac{1}{2}\left(\frac{V''}{h}-\frac{V'h'}{2h^2}\right)^2+\frac{n^2-1}{2}\left(\frac{V'f'}{2hf}\right)^2\,.
\end{align}
The previous equations encode the anomalous dimensions for the decay constants and masses of the Goldstones and the scalar.
 
\subsection{Results for $SU(2)$ and $SU(3)$}

The basis of NLO operators of eq.~(\ref{NLO}) is valid for generic $SU(n)$. However, in the case of $SU(2)$ and $SU(3)$ there are identities that reduce the number of independent NLO operators. It is therefore convenient to write the anomalous dimensions in the reduced, nonredundant basis. 

For $SU(2)$, the number of generic NLO operators can be reduced thanks to the following Cayley-Hamilton identities:
\begin{align}
\langle u_{\mu}u^{\mu}u_{\nu}u^{\nu}\rangle&=\frac{1}{2}\langle u_{\mu}u^{\mu}\rangle^2\nonumber\\
\langle u^{\mu}u^{\nu}u_{\mu}u_{\nu}\rangle&=\langle u_{\mu}u_{\nu}\rangle^2-\frac{1}{2}\langle u_{\mu}u^{\mu}\rangle^2\nonumber\\
\langle u_{\mu}u^{\mu}\chi_+\rangle&=\frac{1}{2}\langle u_{\mu}u^{\mu}\rangle\langle\chi_+\rangle\,.
\end{align}
The anomalous dimensions for the independent operators thus read
\begin{align}\label{su2}
\Gamma_1^{(2)}&=\frac{f^2}{8h^2}{\cal{B}}^2+\frac{1}{12}a^2-\frac{1}{6}a+\frac{1}{12}; &\Gamma_2^{(2)}&=\frac{a^2}{6}-\frac{a}{3}+\frac{1}{6}\nonumber\\
\Gamma_4^{(2)}&=-\frac{f}{8h}{\cal{B}}{\cal{D}}-\frac{g}{4f}(a+b-ba-1); &\Gamma_6^{(2)}&=\frac{1}{32}\left[{\cal{D}}^2+3(1-b)^2\left(\frac{g}{f}\right)^2\right]\nonumber\\
\Gamma_7^{(2)}&=\frac{1}{16}\frac{g^2}{fh}{\cal{C}}^2; &\Gamma_8^{(2)}&=-\frac{1}{8}\frac{g^2}{fh}{\cal{C}}^2\nonumber\\
\Gamma_9^{(2)}&=\frac{1-a}{6}; &\Gamma_{10}^{(2)}&=\frac{a-1}{6}\nonumber\\
\widehat\Gamma_{1}^{(2)}&=\frac{3}{2f_\pi^4}{\cal{B}}^2; &\widehat\Gamma_{2}^{(2)}&=-\frac{1}{f_\pi^2}{\cal{B}}\left(\frac{f}{6h}{\cal{B}}+a-1\right)\nonumber\\
\widehat\Gamma_{3}^{(2)}&=\frac{2}{3f_\pi^2}\frac{f}{h}{\cal{B}}^2; &\widehat\Gamma_{4}^{(2)}&=\frac{3}{4f_\pi^2}\frac{g}{f}{\cal{B}}\left(1-b\right)\nonumber\\
\widehat\Gamma_{5}^{(2)}&=\frac{1}{f_\pi}\frac{g}{h}{\cal{B}}{\cal{C}}; &\widehat\Gamma_{6}^{(2)}&=-\frac{1}{6}\frac{f'}{h}{\cal{B}}\nonumber\\
{\Delta}_{1}^{(2)}&=-\frac{a+1}{12}; &{\Delta}_{2}^{(2)}&=\frac{1}{4}\frac{g^2}{fh}{\cal{C}}^2\,,
\end{align}
where our notation is such that redundant operators are just skipped in the numbering.

The previous results agree with the scalar sector contribution to the one-loop anomalous dimensions of the Higgs-electroweak EFT~\cite{Guo:2015isa,Buchalla:2017jlu,Alonso:2017tdy}. As an additional cross-check of the previous equation, one can apply it to the $SU(2)$ linear sigma model~\cite{GellMann:1960np}, defined by the Lagrangian:
\begin{align}\label{LsM}
{\cal{L}}=\frac{1}{2}\partial_{\mu}S\partial^{\mu}S+\frac{f_\pi^2}{4}\left(1+\frac{S}{f_\pi}\right)^2\langle u_{\mu}u^{\mu}\rangle-gf_\pi\left(1+\frac{S}{f_\pi}\right)(\bar{\psi}_LU\psi_R+{\bar{\psi}}_RU^{\dagger}\psi_L)-V(S)\,,
\end{align}
where the scalar potential $V(S)$ is a function that contains quadratic, cubic and quartic powers of $S$. Since the NLO anomalous dimensions do not depend on the scalar potential, we will leave it in implicit form.

The connection with the generic Lagrangian of eq.~(\ref{LO}) can be done by identifying the spurion matrix $\chi$ with
\begin{align}
\chi_{ji}=-\psi_{Lj}{\bar{\psi}}_{Ri}\,,
\end{align}
and choosing
\begin{align}
f(S)=\left(1+\frac{S}{f_\pi}\right)^2; \qquad g(S)=-\frac{4g}{f_\pi}\left(1+\frac{S}{f_\pi}\right);\qquad h(S)=1\,.
\end{align}
For these particular values of the scalar functions, one obtains
\begin{align}
a&=b=1;\qquad &{\cal{B}}&={\cal{C}}={\cal{D}}=0\,.
\end{align}
One can then easily check that for this model the anomalous dimensions in eq.~(\ref{su2}), with the exception of $\Delta_1$, identically vanish. Since the linear sigma model is a renormalizable theory, this in particular entails that it has no NLO counterterms. Therefore, the cancellation of the anomalous dimensions is not just expected but actually a consistency check. For simplicity, in eq.~(\ref{LsM}) we have set the left- and right-handed sources to zero. Were these to be included and turned into dynamical fields, the operator associated to $\Delta_1$ would get renormalized. Actually, $\Delta_1$ is nothing but the contribution of the Goldstone modes to the beta function of the external sources, if these are promoted to gauge fields. It is therefore a one-loop renormalization of the gauge field kinetic term. 

Let us now move to $SU(3)$. In this case, there is only one Cayley-Hamilton relation to be used:
\begin{align}
\langle u_{\mu}u_{\nu}u^{\mu}u^{\nu}\rangle=\langle u_{\mu}u_{\nu}\rangle^2+\frac{1}{2}\langle u_{\mu}u^{\mu}\rangle^2-2\langle u_{\mu}u^{\mu}u_{\nu}u^{\nu}\rangle\,,
\end{align}
and the set of anomalous dimensions then take the form:
\begin{align}\label{su3ad}
\Gamma_1^{(3)}&=\frac{f^2}{8h^2}{\cal{B}}^2+\frac{17}{24}a^2-\frac{11}{12}a+\frac{3}{32}; &\Gamma_2^{(3)}&=\frac{a^2}{6}-\frac{a}{3}+\frac{3}{16}\nonumber\\
\Gamma_3^{(3)}&=a; &\Gamma_4^{(3)}&=-\frac{g}{f}\frac{1}{4}\left(\frac{f^2}{2gh}{\cal{B}}{\cal{D}}+\frac{8}{3}a\left(1-\frac{21}{16}b\right)-\frac{1-3b}{2}\right)\nonumber\\
\Gamma_5^{(3)}&=\frac{g}{f}\left(\frac{a}{4}+\frac{3}{8}\right); &\Gamma_6^{(3)}&=\frac{1}{32}\left(\frac{g}{f}\right)^2\left(\frac{f^2}{g^2}{\cal{D}}^2+8b^2-\frac{32}{3}b+\frac{22}{9}\right)\nonumber\\
\Gamma_7^{(3)}&=\frac{1}{24}\frac{g^2}{fh}{\cal{C}}^2; &\Gamma_8^{(3)}&=\frac{5}{48}\left(\frac{g}{f}\right)^2-\frac{1}{8}\frac{g^2}{fh}{\cal{C}}^2\nonumber\\
\Gamma_9^{(3)}&=-\left(\frac{a}{6}-\frac{1}{4}\right); &\Gamma_{10}^{(3)}&=\frac{a}{6}-\frac{1}{4}\nonumber\\
\widehat\Gamma_{1}^{(3)}&=\frac{4}{f_\pi^4}{\cal{B}}^2; &\widehat\Gamma_{2}^{(3)}&=-\frac{1}{f_\pi^2}{\cal{B}}\left(\frac{f}{6h}{\cal{B}}+\frac{7}{2}a-\frac{3}{2}\right)\nonumber\\
\widehat\Gamma_{3}^{(3)}&=\frac{2}{3f_\pi^2}\frac{f}{h}{\cal{B}}^2; &\widehat\Gamma_{4}^{(3)}&=\frac{3}{4f_\pi^2}\frac{g}{f}{\cal{B}}\left(\frac{16}{9}-\frac{8}{3}b\right)\nonumber\\
\widehat\Gamma_{5}^{(3)}&=\frac{1}{f_\pi}\frac{g}{h}{\cal{B}}{\cal{C}}; &\widehat\Gamma_{6}^{(3)}&=-\frac{1}{6}\frac{f'}{h}{\cal{B}}\nonumber\\
{\Delta}_{1}^{(3)}&=-\frac{a}{12}-\frac{1}{8}; &{\Delta}_{2}^{(3)}&=\frac{5}{24}\left(\frac{g}{f}\right)^2+\frac{1}{4}\frac{g^2}{fh}{\cal{C}}^2\,.
\end{align}
As a particular case of the previous list, one can examine $SU(3)$ ChPT, i.e. a theory without a scalar field. In this case, $g(\phi)=f(\phi)=1$ and all the functions defined in eqs.~(\ref{rel1}) and (\ref{rel2}) vanish. One ends up with the well-known result~\cite{Gasser:1984gg}
\begin{align}
\Gamma_1&=\frac{3}{32};&\Gamma_2&=\frac{3}{16};&\Gamma_3&=0;&\Gamma_4&=\frac{1}{8};&\Gamma_5&=\frac{3}{8}\nonumber\\
\Gamma_6&=\frac{11}{144};&\Gamma_7&=0;&\Gamma_8&=\frac{5}{48};&\Gamma_9&=\frac{1}{4};&\Gamma_{10}&=-\frac{1}{4}\nonumber\\
{\Delta}_{1}&=-\frac{1}{8};&{\Delta}_{2}&=\frac{5}{24}\,,
\end{align}
together with $\widehat{\Gamma}_j=0$, as expected.
 
\section{Chiral-scale perturbation theory to one loop}
\label{sec:4}

In this section, we will apply our results to one particular example which is not only interesting from a conceptual point of view but also from the phenomenological side.

For quite some time, there has been an active discussion about the nature (and even the existence) of the $f_0(500)$ meson (the so-called $\sigma$ meson) in the hadronic spectrum. Its existence nowadays is out of question~\cite{Caprini:2005zr,Tanabashi:2018oca}, but its nature is still unclear~\cite{Pelaez:2003dy}. 

In ChPT, the sigma is a resonance state and, accordingly, their effects are integrated out in the low-energy effective action. However, it is even lighter than the kaon. This means that in those channels where the sigma can be exchanged, the chiral expansion is only valid at energies below the $\sigma$ threshold. 
	
Chiral-scale perturbation theory~\cite{Crewther:2012wd,Crewther:2013vea,Crewther:2015dpa} (ChSPT) instead treats the $\sigma$ meson as a light degree of freedom alongside the pion multiplet and interprets it as a Goldstone mode of broken scale invariance. One therefore assumes the existence of a nonperturbative infrared fixed point, where the running of $\alpha_{s}$ freezes. Under this assumption, the chiral condensate that spontaneously breaks chiral symmetry also breaks spontaneously scale invariance~\cite{Crewther:1970gx,Ellis:1970yd,Crewther:1971bt,Leung:1989hw}. As a result, the spectrum has a dilaton, which is identified with the $\sigma$ meson. 

The most salient feature that distinguishes ChSPT from $SU(3)$ ChPT is in kaon decays, where the $\sigma$ exchange provides a natural resolution of the $\Delta I=\tfrac{1}{2}$ rule~\cite{Crewther:2012wd,Crewther:2013vea}. This is a tree level effect. However, the fact that the $\sigma$ is considered as a light degree of freedom means that the renormalization-group (RG) evolution of the NLO coefficients differs from that of $SU(3)$ ChPT. One therefore expects differences in both theories at the loop level in the pure strong sector. A detailed comparison between both theories at one loop would require not just the anomalous dimensions but the full renormalization, including finite terms, together with an exhaustive study of the associated phenomenology. This is clearly beyond the scope of the present paper. We will content ourselves here with some qualitative considerations.   

There is a well-defined, rather simple procedure of how to write down theories of Goldstone modes coupled to a dilaton~\cite{Ellis:1970yd,Zumino:1970tu}. Let us first consider spontaneous breaking only. The chiral-invariant sector of the Lagrangian gets upgraded to scale invariance by introducing a chiral singlet scale compensator ${\hat{\chi}}$, transforming linearly under scale transformations with scaling dimension 1. Infinitesimally,
\begin{align}
\hat{\chi}(x)\longrightarrow \hat{\chi}(x)+\alpha(1+x^{\mu}\partial_{\mu})\hat{\chi}(x)\,,
\end{align} 
with $\alpha$ a real parameter. We will assume the compensator to be dimensionless (i.e. to carry no canonical dimensions). An operator ${\cal{O}}$ with scaling dimension $d_{\cal{O}}$ is made scale invariant by attaching $(4-d_{\cal{O}})$ powers of the compensator to it. For Goldstone fields and derivatives, the scaling dimension is the chiral dimension. The leading-order Lagrangian is thus simply 
\begin{align}
{\cal{L}}&=\frac{f_{\pi}^2}{4}\langle u_{\mu}u^{\mu}\rangle{\hat{\chi}}^2+\frac{f_{\sigma}^2}{2}\partial_{\mu}{\hat{\chi}}\partial^{\mu}{\hat{\chi}}\,,
\end{align}
where, at this stage, $f_{\sigma}$ is introduced for dimensional reasons. The scale compensator is dynamical, such that it can be identified with the dilaton. From the previous equation and our discussion of power counting in Section~\ref{sec:2}, it is clear that the scale compensator, as any scalar field, has chiral dimension $0$. 

Let us consider now the sources of explicit breaking. Chiral symmetry is explicitly broken by the quark masses, while scale invariance is broken by both the quark masses and the gluonic anomaly. At leading order, the only possible operators of explicit breaking are then
\begin{align}\label{potential}
{\cal{L}}&=\frac{f_{\pi}^2}{4}\langle\chi_+\rangle{\hat{\chi}}^{\gamma}+c_1{\hat{\chi}}^{\delta}-c_2{\hat{\chi}}^4\,,
\end{align}
which correspond to a potential for the dilaton and chiral Goldstone modes. The parameters $c_1$, $c_2$, $\gamma$ and $\delta$ can be determined through matching with QCD, as we will do below. The last term above, which looks superficially scale invariant, has to be present to ensure a consistent dilaton potential (see the comments in~\cite{Zumino:1970tu,Leung:1989hw,Cata:2018wzl}). Without explicit breaking, the dilaton and the chiral Goldstones should be exactly massless and the potential should vanish altogether, as required by Goldstone's theorem. Since the operator ${\hat{\chi}}^4$ is scale invariant, it follows that the coefficient $c_2$ has to break scale symmetry explicitly (see eq.~(\ref{Vcoeff}) below).
  
The coefficients $\gamma$ and $\delta$ can be fixed in terms of QCD parameters by matching the trace anomaly at different scales. The energy-momentum tensor is defined as 
\begin{align}
\theta_{\mu\nu}&=2\frac{\delta}{\delta g^{\mu\nu}}(\sqrt{-g}{\cal{L}})\Bigg|_{g_{\mu\nu}=\eta_{\mu\nu}}=-\eta_{\mu\nu}{\cal{L}}+2\frac{\partial {\cal{L}}}{\partial g^{\mu\nu}}\Bigg|_{g_{\mu\nu}=\eta_{\mu\nu}}\,,
\end{align}
where $g$ is the determinant of the metric tensor $g_{\mu\nu}$ and $\eta_{\mu\nu}$ the Minkowski background. Its trace is generically given by 
\begin{align}\label{generic}
\theta^{\mu}_{\,\,\mu}=\sum_{j} (d_{{\cal{O}}_{j}}-4)\,{\cal{O}}_j\,,
\end{align}
where $d_{{\cal{O}}_j}$ is the scaling dimension of the operator ${\cal{O}}_j$. The result in terms of quarks and gluons is given by the well-known result:
\begin{align}\label{trQCD}
\theta^{\mu}_{\,\,\mu}&=\frac{\beta}{4\alpha_s}G_{\mu\nu}^aG^{\mu\nu\,a}+(1+\gamma_m)\sum_q m_q{\bar{q}}q\,,
\end{align}
which indicates that scale breaking is caused by the gluonic anomaly and by the presence of mass terms, which in turn are also corrected by the anomaly via $\gamma_m$, the anomalous dimension of the ${\bar{q}}q$ operator.
  
The counterpart of eq.~(\ref{trQCD}) at low energies requires some additional steps. For scale-invariant theories with scalars the gravitational energy-momentum tensor has to be improved~\cite{Callan:1970ze}. In ChSPT, one finds~\cite{Crewther:2015dpa} 
\begin{align}
\Theta_{\mu\nu}&=\theta_{\mu\nu}+I_{\mu\nu}\,,
\end{align}
with
\begin{align}
I_{\mu\nu}&=\frac{f_{\sigma}^2}{6}(\eta_{\mu\nu}\Box-\partial_{\mu}\partial_{\nu}){\hat{\chi}}^2\,.
\end{align}
With the previous extension, and using the equations of motion, the trace of the energy-momentum tensor can be shown to be proportional to the scale-breaking terms:
\begin{align}
\Theta^{\mu}_{\,\,\mu}&=(\delta-4)c_1{\hat{\chi}}^{\delta}+(\gamma-4)\frac{f_{\pi}^2}{4}\langle\chi_+\rangle {\hat{\chi}}^{\gamma}\,.
\end{align}
Comparing the previous equation with eq.~(\ref{generic}), and using that the anomalous dimension of the operator $G_{\mu\nu}^aG^{\mu\nu\,a}$ at the fixed point $\alpha_{\mathrm{IR}}$ is given by $\beta'\equiv \tfrac{\partial}{\partial \alpha_s}\beta\big|_{\alpha_{\mathrm{IR}}}$~\cite{Grinstein:1988wz,Crewther:2012wd}, one concludes that $\gamma=3-\gamma_m$ and $\delta=4+\beta'$. Since ChSPT is an expansion in the vicinity of a nonperturbative fixed point, $\gamma_m$ and $\beta'$ are nonperturbative quantities, to be determined e.g. via lattice simulations. 

It only remains to fix the coefficients $c_1$ and $c_2$. In order to do that, it is convenient to rewrite the theory in terms of a canonically normalized dilaton field $\sigma$. This can be achieved through the assignment 
\begin{align}
{\hat{\chi}}=e^{\sigma/f_{\sigma}}\,,
\end{align}
where $\sigma$ transforms nonlinearly under scale transformations. Infinitesimally,
\begin{align}
\sigma(x)\longrightarrow \sigma(x)+\alpha x^{\mu}\partial_{\mu}\sigma(x)+\alpha f_{\sigma}\,.
\end{align}
The expressions for the potential coefficients follow from minimizing the potential in eq.~(\ref{potential}) and matching its curvature with respect to $\sigma$ to the dilaton mass $m_{\sigma}$. From these two conditions, one finds 
\begin{align}\label{Vcoeff}
c_1&=-\frac{1}{\beta'(4+\beta')}f_{\sigma}^2m_{\sigma}^2+\frac{(3-\gamma_m)(1+\gamma_m)}{2\beta'(4+\beta')}f_{\pi}^2(m_{\pi^+}^2+m_{K^+}^2+m_{K^0}^2)\nonumber\\
c_2&=-\frac{1}{4\beta'}f_{\sigma}^2m_{\sigma}^2+\frac{(3-\gamma_m)(1+\gamma_m+\beta')}{8\beta'}f_{\pi}^2(m_{\pi^+}^2+m_{K^+}^2+m_{K^0}^2)\,.
\end{align} 
As expected, the potential coefficients are proportional to the dilaton and chiral Goldstone masses, i.e. to the sources of explicit breaking. As a result, $[c_j]_{\chi}=2$, in agreement with the chiral counting for the leading-order interactions. 

The final result for the Lagrangian at LO can be written in closed form as
\begin{align}
{\cal{L}}&=\left[\frac{f_{\pi}^2}{4}\langle u_{\mu}u^{\mu}\rangle+\frac{1}{2}\partial_{\mu}\sigma\partial^{\mu}\sigma\right]e^{2\sigma/f_{\sigma}}+\frac{f_{\pi}^2}{4}\langle\chi_+\rangle e^{(3-\gamma_m)\sigma/f_{\sigma}}+c_1e^{(4+\beta')\sigma/f_{\sigma}}-c_2e^{4\sigma/f_{\sigma}}\,,
\end{align}  
with $c_1$ and $c_2$ given in eq.~(\ref{Vcoeff}). This Lagrangian agrees with the one given in refs.~\cite{Crewther:2012wd,Crewther:2013vea,Crewther:2015dpa} once a careful distinction between LO and NLO terms is done. Matching it to the generic parametrization of eq.~(\ref{LO}), the scalar functions defined there take the form
\begin{align}
f(\sigma)=h(\sigma)=e^{2\sigma/f_{\sigma}};\qquad g(\sigma)&=e^{(3-\gamma_m)\sigma/f_{\sigma}};\qquad V(\sigma)=c_2e^{4\sigma/f_{\sigma}}-c_1e^{(4+\beta')\sigma/f_{\sigma}}\,.
\end{align}
One can now use eq.~(\ref{su3ad}) to determine the anomalous dimensions for ChSPT. Defining $\gamma\equiv 3-\gamma_m$ for compactness, one finds
\begin{align}
a=\frac{f_\pi^2}{f_{\sigma}^2}\equiv \xi^2;\quad b=\gamma \xi^2;\quad {\cal{B}}=0;\quad {\cal{C}}=(1-\gamma)\xi;\quad {\cal{D}}=\gamma(\gamma-1)\xi^2 e^{(\gamma-2)\sigma/f_{\sigma}}\,,
\end{align}
and the anomalous dimensions take the form
\begin{align}\label{ChSAD}
\Gamma_1^{(\sigma)}&=\frac{3}{32}-\frac{11}{12}\xi^2+\frac{17}{24}\xi^4; &\Gamma_2^{(\sigma)}&=\frac{3}{16}-\frac{1}{3}\xi^2+\frac{1}{6}\xi^4\nonumber\\
\Gamma_3^{(\sigma)}&=\xi^2; &\Gamma_4^{(\sigma)}&=\left(\frac{1}{8}-\frac{3\gamma\xi^2}{8}-\frac{2}{3}\xi^2+\frac{7}{8}\gamma\xi^4\right)e^{(\gamma-2)\sigma/f_{\sigma}}\nonumber\\
\Gamma_5^{(\sigma)}&=\left(\frac{3}{8}+\frac{1}{4}\xi^2\right)e^{(\gamma-2)\sigma/f_{\sigma}}; &\Gamma_6^{(\sigma)}&=\left(\frac{11}{144}-\frac{1}{3}\gamma\xi^2+\frac{1}{4}\gamma^2\xi^4+\frac{\gamma^2(\gamma-1)^2}{32}\xi^4\right)e^{2(\gamma-2)\sigma/f_{\sigma}}\nonumber\\
\Gamma_7^{(\sigma)}&=\frac{1}{24}\xi^2(1-\gamma)^2e^{2(\gamma-2)\sigma/f_{\sigma}}; &\Gamma_8^{(\sigma)}&=\left(\frac{5}{48}-\frac{1}{8}\xi^2(1-\gamma)^2\right)e^{2(\gamma-2)\sigma/f_{\sigma}}\nonumber\\
\Gamma_9^{(\sigma)}&=\frac{1}{4}-\frac{1}{6}\xi^2; &\Gamma_{10}^{(\sigma)}&=-\frac{1}{4}+\frac{1}{6}\xi^2\nonumber\\
\widehat\Gamma_{j}^{(\sigma)}&=0,\quad j=1,\dots,6 &&\nonumber\\
{\Delta}_{1}^{(\sigma)}&=-\frac{1}{8}-\frac{1}{12}\xi^2; &{\Delta}_{2}^{(\sigma)}&=\left(\frac{5}{24}+\frac{1}{4}\xi^2(1-\gamma)^2\right)e^{2(\gamma-2)\sigma/f_{\sigma}}\,.
\end{align}
Notice that the divergences to be absorbed at NLO only depend on $\xi$ and $\gamma$. In contrast, the renormalization of the LO coefficients is also sensitive to the form of the potential. Therefore, the renormalization of the decay constants and masses also depends on $\beta'$ and $m_{\sigma}$.

From the results of eq.~(\ref{ChSAD}), one can list a number of observations: 
\begin{itemize}
\item In the chiral limit, $\chi\to 0$, NLO operators have already scaling dimension 4. Accordingly, no dilaton coupling is needed and the NLO anomalous functions are pure numbers. 
\item The operators proportional to $G_j$ do not get renormalized at one loop and thus are $\mu$-independent. No one-loop divergences take the form of these operators.
\item A dilaton dependence only shows up in operators of explicit chiral breaking. Interestingly, however, the dilaton dependence cancels altogether if $\gamma=2$, corresponding to $\gamma_m=1$. For this value, the chiral-breaking operators become scale invariant. In particular, this means that the scaling dimension of $\chi_+$ is 2. We therefore find that the scaling dimension of any chiral building block is actually its chiral dimension. 
\end{itemize} 
With the previous insights, one can write down the NLO Lagrangian for ChSPT. As we discussed, the precise $\sigma$-dependence of the counterterms in eq.~(\ref{ChSAD}) is a consequence of a symmetry principle. It should therefore not only apply to the divergent pieces but also to the finite NLO terms. One can therefore conclude that the NLO Lagrangian is of the form  
\begin{align}\label{NLOChSPT}
{\cal{L}}_4&=L_1\langle u_{\mu}u^{\mu}\rangle^2+L_2\langle u^{\mu}u^{\nu}\rangle\langle u_{\mu}u_{\nu}\rangle+L_3\langle u_{\mu}u^{\mu}u_{\nu}u^{\nu}\rangle\nonumber\\
&+L_4(\sigma)\langle u_{\mu}u^{\mu}\rangle\langle\chi_+\rangle+L_5(\sigma)\langle u_{\mu}u^{\mu}\chi_+\rangle+L_6(\sigma)\langle\chi_+\rangle^2+L_7(\sigma)\langle\chi_-\rangle^2+\frac{1}{2}L_8(\sigma)\langle\chi_+^2+\chi_-^2\rangle\nonumber\\
&-iL_9\langle f_+^{\mu\nu}u_{\mu}u_{\nu}\rangle+\frac{1}{4}L_{10}\langle (f_+^{\mu\nu})^2-(f_-^{\mu\nu})^2\rangle+H_1\langle (f_R^{\mu\nu})^2+(f_L^{\mu\nu})^2\rangle+H_2(\sigma)\langle\chi^{\dagger}\chi\rangle\nonumber\\
&+G_1\partial_{\mu}\sigma\partial^{\mu}\sigma\partial_{\nu}\sigma\partial^{\nu}\sigma+G_2\langle u_{\mu}u^{\mu}\rangle \partial_{\nu}\sigma\partial^{\nu}\sigma+G_3\langle u_{\mu}u_{\nu}\rangle \partial^{\mu}\sigma\partial^{\nu}\sigma\nonumber\\
&+G_4\langle\chi_+\rangle \partial_{\mu}\sigma\partial^{\mu}\sigma+\frac{i}{2}G_5\langle u_{\mu}\chi_-\rangle\partial^{\mu}\sigma+G_6\langle u_{\mu}f_-^{\mu\nu}\rangle\partial_{\nu}\sigma\,,
\end{align}    
where the $\sigma$ dependence of the $L_i(\sigma)$ functions can be read from eq.~(\ref{ChSAD}). 

The NLO basis above differs from the one of ref.~\cite{Li:2016uzn}. In particular, in ref.~\cite{Li:2016uzn} (i) the operator with coefficient $G_5$ is absent and (ii) when $\gamma_m=1$, the set of NLO operators does not reduce to the expected scale-invariant limit, in which all the NLO coefficients should be constant. The starting Lagrangian of ref.~\cite{Li:2016uzn} is taken from refs.~\cite{Crewther:2012wd,Crewther:2013vea,Crewther:2015dpa}, where LO and NLO terms were not carefully sorted out. However, this difference with our work cannot be the origin of the discrepancies noted above.

We will finish this section with some comments on the comparison of $SU(3)$ ChPT and ChSPT on the basis of our results above. The most direct way of testing ChSPT is to search for an infrared fixed point, where $\alpha_s$ freezes and $m_{\sigma}$ vanishes. This can in principle be done with lattice simulations, but in practice it can be difficult to identify fixed points. As we pointed out at the beginning of this section, the fact that the $\sigma$ is included as a light degree of freedom means that the RG evolution of the NLO coefficients gets modified, as compared to the predictions of ChPT. Thus, an alternative to the search for fixed points is to test the RG evolution of the parameters of $SU(3)$ ChPT, e.g. by determining them at different scales and extracting the anomalous dimension from the slope of the logarithmic plot. For instance, for the NLO coefficients, one would need to fit the functions
\begin{align}
\mu\frac{d}{d\mu}L_i^r(\mu)&=-\frac{\Gamma_i}{16\pi^2}\,.
\end{align}
Given a target precision for the $L_i^r(\mu)$, one can estimate the window of scales to test ChPT against ChSPT. Assuming a measurement at a reference scale $\mu_0$, at a scale $\mu$ one finds
\begin{align}
L_i^{(\sigma)}(\mu)=L_i^{(\chi)}(\mu)+\frac{\Gamma_i^{(\sigma)}-\Gamma_i^{(\chi)}}{16\pi^2}\log\frac{\mu_0}{\mu}\,,
\end{align}
where $L_i^{(\sigma)}(\mu)$ and $L_i^{(\chi)}(\mu)$ are the renormalized coefficients to be found with the ChSPT and ChPT RG evolution, respectively. 

Notice that for operators without chiral mass insertions, the previous equation simply depends on the ratio of decay constants
\begin{align}
\xi=\frac{f_{\pi}}{f_{\sigma}}\,.
\end{align}
As an example, consider
\begin{align}
L_3^{(\sigma)}(\mu)&=L_3^{(\chi)}(\mu)+\frac{\xi^2}{16\pi^2}\log\frac{\mu_0}{\mu}\nonumber\\
L_{10}^{(\sigma)}(\mu)&=L_{10}^{(\chi)}(\mu)+\frac{\xi^2}{96\pi^2}\log\frac{\mu_0}{\mu}\,.
\end{align} 
Since in ChSPT the chiral condensate is responsible for both spontaneous scale and chiral breaking, $\xi$ should be an ${\cal{O}}(1)$ parameter. Taking $\xi=1$ as a benchmark point, and given an expected precision of $10^{-4}$ at the new scale, it would be enough to deviate a $2\%$ and a $10\%$ from the reference scale $\mu_0$, respectively, to be able to distinguish both scenarios. In practice, a $10^{-4}$ precision corresponds roughly to a $10\%$ uncertainty in the values of the $L_i$. This is rather realistic, given that lattice simulations have already this precision~\cite{Aoki:2019cca}.   
 
\section{Conclusions}
\label{sec:5}
In this paper, we have analyzed the divergent one-loop structure of chiral $SU(n)$ theories extended with a generic scalar singlet $\phi$. The addition of the scalar field generates 6 new operators at NLO, on top of the usual 13 in chiral $SU(n)$ theories. They all get dressed with arbitrary, non-derivative functions of $\phi$, whose precise form depends on the nature of the scalar singlet.

Using the background field method and the heat kernel expansion, we have isolated the one-loop divergences from the effective action and listed the resulting anomalous dimensions for generic $n$, taking into account the reductions that apply for $n=2$ and $n=3$. The resulting master formula has been cross-checked against the $SU(2)$ linear sigma model and $SU(3)$ chiral perturbation theory. 

Our main results can be summarized as follows:
\begin{itemize}
\item The combination $L_9+L_{10}$, which in ChPT is scale-independent, is actually scale-independent also in the presence of generic light scalars. $L_7$ instead acquires a renormalization due to mixed (Goldstone and scalar) one-loop diagrams.
\item Unlike in pure chiral theories, the presence of a scalar potential induces a renormalization of the leading-order parameters of the theory. However, the potential does not participate in the renormalization of the NLO operators.
\item In theories with scale symmetry, whether the symmetry is manifest or hidden is contained in the scalar potential, and thus both scenarios cannot be distinguished from the RG running of the NLO coefficients. However, this distinction is relevant for the scale dependence of the leading-order coefficients, in particular for the scale evolution of the pion and dilaton decay constants and masses. 
\item We have applied our results to chiral-scale perturbation theory (ChSPT), an alternative to $SU(3)$ ChPT where the $f_0(500)$ is interpreted as a pseudo-dilaton of spontaneous scale symmetry breaking. We have determined the NLO operators and their associated anomalous dimensions.
\item In order to test ChSPT with lattice simulations, the exploration of the RG evolution of the NLO coefficients due to virtual light scalar loops can be a rather efficient alternative e.g. to searching for an infrared fixed point. Given the current precision in lattice studies, one can even aim at extracting meaningful bounds on $f_{\sigma}$ or $\gamma_m$.  
\end{itemize}     
The complete one-loop renormalization of a $SU(n)$ chiral effective theory with a generic scalar singlet, including also finite terms, is deferred to a future publication. 

\section*{Acknowledgements}
We would like to thank Gerhard Buchalla for useful discussions at the different stages of this work and valuable comments on the manuscript. The work of O.~C. is supported by the Deutsche Forschungsgemeinschaft (DFG FOR 1873). C.~M. is supported in part by the Deutsche Forschungsgemeinschaft (DFG) under grant BU 1391/2-2 (project number 261324988) and the DFG Cluster Exc 2094 ''ORIGINS''.



\begin{thebibliography}{99}

\bibitem{Aoki:2014oha} 
  Y.~Aoki {\it et al.} [LatKMI Collaboration],
  Phys.\ Rev.\ D {\bf 89}, 111502 (2014)
  [arXiv:1403.5000 [hep-lat]].
		
\bibitem{Fodor:2014pqa} 
  Z.~Fodor, K.~Holland, J.~Kuti, D.~Nogradi and C.~H.~Wong,
  PoS LATTICE {\bf 2013}, 062 (2014)
  [arXiv:1401.2176 [hep-lat]].

\bibitem{Aoki:2013zsa} 
  Y.~Aoki {\it et al.} [LatKMI Collaboration],
  Phys.\ Rev.\ Lett.\  {\bf 111}, no. 16, 162001 (2013)
  [arXiv:1305.6006 [hep-lat]].
	
\bibitem{Appelquist:2018yqe} 
  T.~Appelquist {\it et al.} [Lattice Strong Dynamics Collaboration],
  Phys.\ Rev.\ D {\bf 99}, no. 1, 014509 (2019)
  [arXiv:1807.08411 [hep-lat]].

\bibitem{Appelquist:2018tyt} 
  T.~Appelquist {\it et al.} [LSD Collaboration],
  Phys.\ Rev.\ D {\bf 98}, no. 11, 114510 (2018)
  [arXiv:1809.02624 [hep-ph]].
	
\bibitem{Appelquist:2010gy} 
  T.~Appelquist and Y.~Bai,
  Phys.\ Rev.\ D {\bf 82}, 071701 (2010)
  [arXiv:1006.4375 [hep-ph]].

\bibitem{Cata:2018wzl} 
  O.~Cat\`a, R.~J.~Crewther and L.~C.~Tunstall,
  Phys.\ Rev.\ D {\bf 100}, no. 9, 095007 (2019)
  [arXiv:1803.08513 [hep-ph]].

\bibitem{Callan:1969sn} 
  C.~G.~Callan, Jr., S.~R.~Coleman, J.~Wess and B.~Zumino,
  Phys.\ Rev.\  {\bf 177}, 2247 (1969).
	
\bibitem{Soto:2011ap} 
  J.~Soto, P.~Talavera and J.~Tarrus,
  Nucl.\ Phys.\ B {\bf 866}, 270 (2013)
  [arXiv:1110.6156 [hep-ph]].				
	
\bibitem{Golterman:2016lsd} 
  M.~Golterman and Y.~Shamir,
  Phys.\ Rev.\ D {\bf 94}, no. 5, 054502 (2016)
  [arXiv:1603.04575 [hep-ph]].

\bibitem{Hansen:2016fri} 
  M.~Hansen, K.~Langaeble and F.~Sannino,
  Phys.\ Rev.\ D {\bf 95}, no. 3, 036005 (2017)
  [arXiv:1610.02904 [hep-ph]].
	
\bibitem{Hansen:2018gck} 
  M.~Hansen, K.~Langaeble and F.~Sannino,
  PoS Confinement {\bf 2018}, 222 (2019)
  [arXiv:1810.11993 [hep-ph]].

\bibitem{Abbott:1981ke}
  L.~F.~Abbott,
  Acta Phys.\ Polon.\ B {\bf 13} (1982) 33.

\bibitem{DeWitt:1967yk} 
  B.~S.~DeWitt,
  Phys.\ Rev.\  {\bf 160}, 1113 (1967).
  Phys.\ Rev.\  {\bf 162}, 1195 (1967).
 
\bibitem{Crewther:2012wd} 
  R.~J.~Crewther and L.~C.~Tunstall,
  arXiv:1203.1321 [hep-ph].

\bibitem{Crewther:2013vea}
  R.~J.~Crewther and L.~C.~Tunstall,
  Phys.\ Rev.\ D {\bf 91}, no. 3, 034016 (2015)
  [arXiv:1312.3319 [hep-ph]].
	
\bibitem{Crewther:2015dpa} 
  R.~J.~Crewther and L.~C.~Tunstall,
  PoS CD {\bf 15}, 132 (2015)
  [arXiv:1510.01322 [hep-ph]].			

\bibitem{Buchalla:2012qq} 
  G.~Buchalla and O.~Cata,
  JHEP {\bf 1207}, 101 (2012)
  [arXiv:1203.6510 [hep-ph]].
	
\bibitem{Buchalla:2013rka}
  G.~Buchalla, O.~Cat\`a and C.~Krause,
  Nucl.\ Phys.\ B {\bf 880}, 552 (2014)
  [arXiv:1307.5017 [hep-ph]].

\bibitem{Buchalla:2013eza}
  G.~Buchalla, O.~Cat\`a and C.~Krause,
  Phys.\ Lett.\ B {\bf 731}, 80 (2014)
  [arXiv:1312.5624 [hep-ph]].

\bibitem{Buchalla:2016sop} 
  G.~Buchalla, O.~Cat\`a, A.~Celis and C.~Krause,
  arXiv:1603.03062 [hep-ph].

\bibitem{Urech:1994hd} 
  R.~Urech,
  Nucl.\ Phys.\ B {\bf 433}, 234 (1995)
  [hep-ph/9405341].

\bibitem{Gasser:1983yg}
  J.~Gasser and H.~Leutwyler,
  Annals Phys.\  {\bf 158}, 142 (1984).
	
\bibitem{tHooft:1973bhk}
  G.~'t Hooft,
  Nucl.\ Phys.\ B {\bf 62} (1973) 444.

\bibitem{Georgi:1991ch} 
  H.~Georgi,
  Nucl.\ Phys.\ B {\bf 361}, 339 (1991).

\bibitem{Guo:2015isa}
  F.~K.~Guo, P.~Ruiz-Femen\'ia and J.~J.~Sanz-Cillero,
  Phys.\ Rev.\ D {\bf 92} (2015) 074005
  [arXiv:1506.04204 [hep-ph]].
	
\bibitem{Buchalla:2017jlu}
  G.~Buchalla, O.~Cat\`a, A.~Celis, M.~Knecht and C.~Krause,
  Nucl.\ Phys.\ B {\bf 928}, 93 (2018)
  [arXiv:1710.06412 [hep-ph]].

\bibitem{Alonso:2017tdy} 
  R.~Alonso, K.~Kanshin and S.~Saa,
  Phys.\ Rev.\ D {\bf 97}, no. 3, 035010 (2018)
  [arXiv:1710.06848 [hep-ph]].
	
\bibitem{GellMann:1960np} 
  M.~Gell-Mann and M.~Levy,
  Nuovo Cim.\  {\bf 16}, 705 (1960).
				
\bibitem{Gasser:1984gg}
  J.~Gasser and H.~Leutwyler,
  Nucl.\ Phys.\ B {\bf 250}, 465 (1985).

\bibitem{Caprini:2005zr} 
  I.~Caprini, G.~Colangelo and H.~Leutwyler,
  Phys.\ Rev.\ Lett.\  {\bf 96}, 132001 (2006)
  [hep-ph/0512364].

\bibitem{Tanabashi:2018oca} 
  M.~Tanabashi {\it et al.} [Particle Data Group],
  Phys.\ Rev.\ D {\bf 98}, no. 3, 030001 (2018).

\bibitem{Pelaez:2003dy} 
  J.~R.~Pelaez,
  Phys.\ Rev.\ Lett.\  {\bf 92}, 102001 (2004)
  [hep-ph/0309292].

\bibitem{Crewther:1970gx} 
  R.~J.~Crewther,
  Phys.\ Lett.\  {\bf 33B}, 305 (1970).
 
\bibitem{Ellis:1970yd}
  J.~R.~Ellis,
  Nucl.\ Phys.\ B {\bf 22} (1970) 478
   Erratum: [Nucl.\ Phys.\ B {\bf 25} (1971) 639].
		
\bibitem{Crewther:1971bt} 
  R.~J.~Crewther,
  Phys.\ Rev.\ D {\bf 3}, 3152 (1971)
  Erratum: [Phys.\ Rev.\ D {\bf 4}, 3814 (1971)].

\bibitem{Leung:1989hw} 
  C.~N.~Leung, S.~T.~Love and W.~A.~Bardeen,
  Nucl.\ Phys.\ B {\bf 323}, 493 (1989).

\bibitem{Zumino:1970tu} 
  B.~Zumino,
  Lectures on Elementary Particles and Quantum Field Theory v.2, Cambridge, Mass.: Brandeis Univ., pp. 437-500.

\bibitem{Callan:1970ze} 
  C.~G.~Callan, Jr., S.~R.~Coleman and R.~Jackiw,
  Annals Phys.\  {\bf 59}, 42 (1970).
					
\bibitem{Grinstein:1988wz} 
  B.~Grinstein and L.~Randall,
  Phys.\ Lett.\ B {\bf 217}, 335 (1989).

\bibitem{Li:2016uzn} 
  Y.~L.~Li, Y.~L.~Ma and M.~Rho,
  Phys.\ Rev.\ D {\bf 95}, no. 11, 114011 (2017)
  [arXiv:1609.07014 [hep-ph]].
		
\bibitem{Aoki:2019cca} 
  S.~Aoki {\it et al.} [Flavour Lattice Averaging Group],
  arXiv:1902.08191 [hep-lat].
			
\end{thebibliography}
\end{document}